\documentclass[runningheads]{llncs}
\usepackage{eso-pic}
\definecolor{lightgray}{gray}{0.5}

\usepackage{graphicx}
\usepackage{multirow}
\usepackage{amsmath,amssymb,amsfonts}%
\usepackage{mathrsfs}
\usepackage{eso-pic}
\definecolor{lightgray}{gray}{0.5}

\usepackage{xcolor}
\usepackage{textcomp}
\usepackage[utf8]{inputenc}
\usepackage{newunicodechar}
\newunicodechar{Ω}{\Omega}

\usepackage{manyfoot}
\usepackage{booktabs}
\usepackage{algorithm}
\usepackage{algorithmicx}
\usepackage{algpseudocode}
\usepackage{listings}
\usepackage{graphicx}
\usepackage{subcaption}
\usepackage{siunitx}
\usepackage{mathrsfs}
\usepackage{eso-pic}
\definecolor{lightgray}{gray}{0.5}

\begin{document}

\AddToShipoutPictureBG*{
  \AtPageUpperLeft{
    \raisebox{-1.5cm}{
      \makebox[\paperwidth]{
        \begin{minipage}{0.9\paperwidth}
          \centering\footnotesize
          \textcolor{lightgray}{This is an open-access, author-archived version of a manuscript published in ApplePies 2025 Conference.}
        \end{minipage}
      }
    }
  }
}
\AddToShipoutPictureBG*{
  \AtPageLowerLeft{
    \raisebox{1cm}{
      \makebox[\paperwidth]{
        \begin{minipage}{0.9\paperwidth}
          \centering\footnotesize
          \textcolor{lightgray}{©2025 Authors \& Springer. This is the author's version of the work. It is posted here for your personal use. Not for redistribution. \\ 
          The definitive Version of Record is published in the Proceedings of the ApplePies 2025 Conference.}
        \end{minipage}
      }
    }
  }
}

\title{Solar Panel-based Visible Light Communication for Batteryless Systems}
\titlerunning{Solar Panel-based VLC for Batteryless Systems}
\author{Juan F. Gutiérrez \inst{1,2} \and
Nhung Nguyen \inst{1} \and
Jesús M. Quintero\inst{2} \and
Andres Gomez \inst{1}
}
\authorrunning{J.F. Gutierrez et al.}
\institute{Institut für Datentechnik und Kommunikationsnetze, Technische Universität Braunschweig, Hans-Sommer-Str. 66, Braunschweig, 38106, Germany
\email{juan-felipe.gutierrez-gomez@tu-braunschweig.de}\\
\and
School of Engineering, Department of Electrical and Electronic
Engineering, Universidad Nacional de Colombia, Ave Cra 30 No. 45-3, Bogotá, 111321, Colombia
}
\maketitle              

\begin{abstract}
This paper presents a batteryless wireless communication node for the Internet of Things, powered entirely by ambient light and capable of receiving data through visible light communication. A solar panel serves dual functions as an energy harvester and an optical antenna, capturing modulated signals from LED light sources. A lightweight analog front-end filters and digitizes the signals for an 8-bit low-power processor, which manages the system's operational states based on stored energy levels. The main processor is selectively activated to minimize energy consumption. Data reception is synchronized with the harvester’s open-circuit phase, reducing interference and improving signal quality. The prototype reliably decodes 32-bit VLC frames at \SI{800}{\hertz}, consuming less than \SI{2.8}{\milli\joule}, and maintains sleep-mode power below 30\,$\mu$W.
\end{abstract}


\section{Introduction}\label{sec1}

In modern IoT systems, low-power embedded devices are critical to enabling long-term autonomous operation in environments where frequent maintenance or a reliable power supply is not feasible. One notable advancement is the development of batteryless sensing nodes, which harvest ambient energy such as solar energy harvesting, RF energy harvesting, or mechanical vibrations to power their sensing and communication tasks~\cite{2}. 
The harvester not only commences a cold start with minimum input power but also regulates voltage and handles energy storage, ensuring stable operation under sporadically available energy conditions.

One of the main obstacles in energy-constrained embedded systems is data reception without significantly increasing energy consumption. Traditional radio frequency (RF) receivers frequently require complicated analog front-ends and high-frequency clocks, resulting in significant standby and active power overhead~\cite{14}. In contrast, Visible Light Communication (VLC) provides a low-power alternative, with simple photodiodes and low-complexity digital circuits used to receive~\cite{15}. In addition, it allows the solar panel to function as both an energy harvester and a receiver, reducing additional hardware and power expenditures.However, the entire potential of VLC reception is still underused. With reliable VLC-based data reception, embedded devices may perform operations such as remote reconfiguration, firmware updates, all without relying on power-hungry radios. These characteristics are especially useful for maintenance-free IoT deployments in remote places. 

Few studies have looked at batteryless IoT devices, MiroCard and DPP3E being notable examples. The active period of MiroCard is limited to a few milliseconds due to the usage of two \SI{47}{\micro\farad} capacitors and communicates with BLE, Zigbee, 6LoWPAN and RF4CE through a \SI{2.4}{\giga\hertz} inverted F antenna. DPP3E provides higher energy autonomy using configurable supercapacitors, uses an Apollo3 Blue Plus MCU with BLE and an AMCA31 chip antenna, as well as an STM32L433 and Semtech SX1262 LoRa transceiver for long-range communication~\cite{17,18}. Another recent research presented batteryless sensor nodes powered by indoor light harvesting, using either BLE or combined VLC/IR communication. The VLC/IR approach, despite higher energy consumption than BLE, but higher accuracy data retention~\cite{10817258}. To enhance adaptability, we propose a hybrid architecture combining BLE for frequent updates with VLC for robust, energy-efficient downlink control, achieving responsiveness and efficiency.

This paper presents a batteryless BLE node powered exclusively by a solar-based energy harvester. VLC signals received through the solar panel are detected by a low-power 8-bit processor via GPIO, which monitors channel activity and selectively wakes the main processor. Depending on channel conditions, the main processor executes either a BLE or VLC decoding application.
Our prototype, implemented using a Texas Instruments CC1352P1 LaunchPad with Zephyr RTOS, successfully demonstrates reliable reception and decoding of 32-bit VLC frames at \SI{800}{\hertz} of optical clock. The reception of VLC data is carried out during the open-circuit state of the energy harvester, which enhances signal quality by minimizing interference. 

\section{System design and Implementation}\label{sec3}

\subsection{System Design}\label{subsec2}

\begin{figure}[htbp]
  \centering
  \resizebox{0.99\textwidth}{!}{
    \begin{minipage}{\textwidth}
      \begin{subfigure}{0.55\textwidth}
        \includegraphics[width=\textwidth]{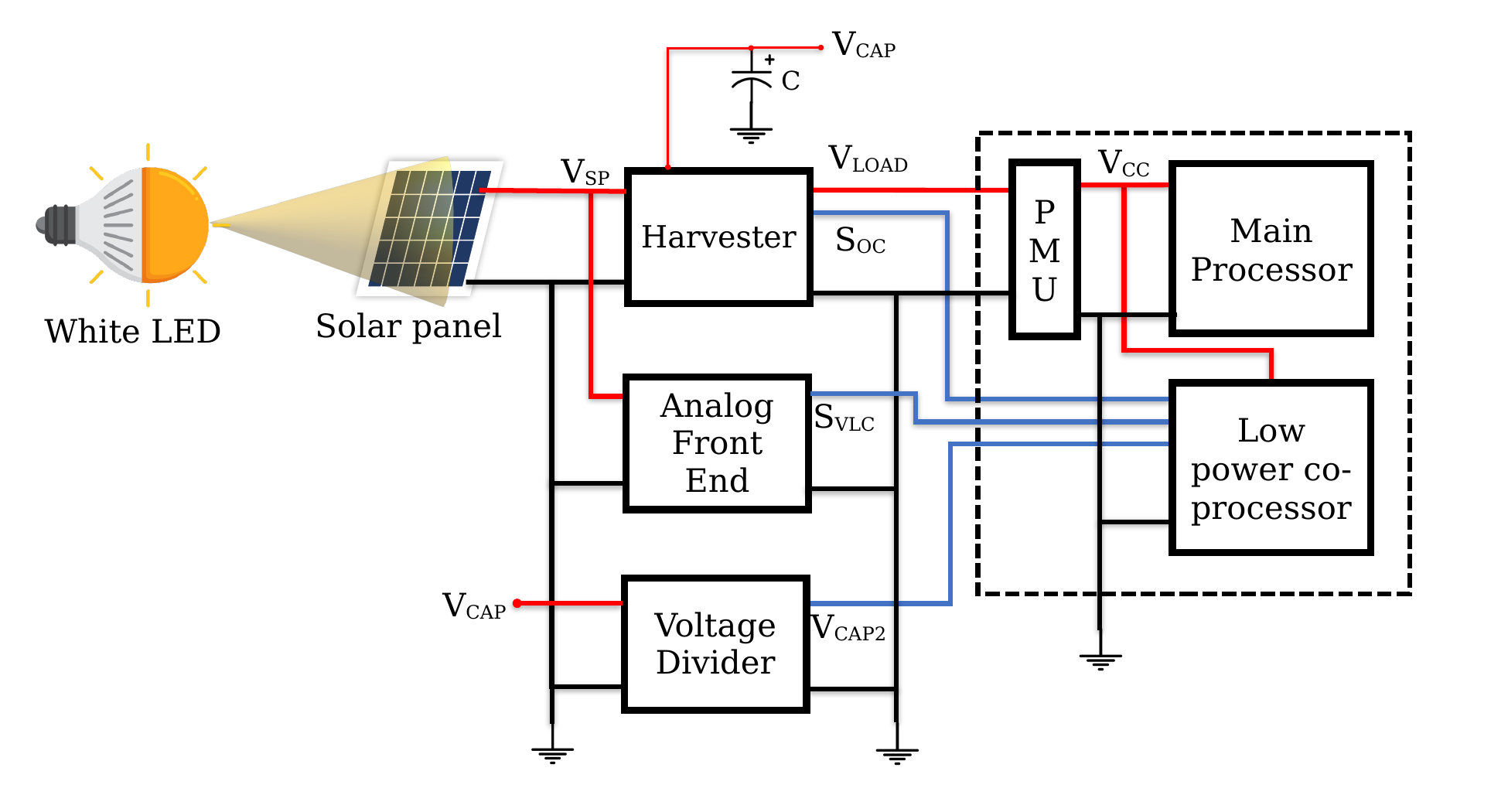}
        \caption{Components of battery-less VLC}
        \label{fig:vlc_system}
      \end{subfigure}
      \hfill
      \begin{subfigure}{0.34\textwidth}
        \includegraphics[width=\textwidth]{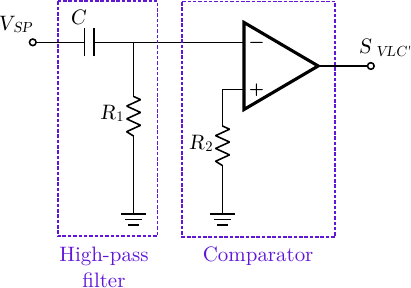}
        \caption{Analog front-end circuit}
        \label{fig:vlc_analog_front_end}
      \end{subfigure}
    \end{minipage}
  }
  \caption{Batteryless VLC Node Architecture and Analog Front-End Design}
  \label{fig:vlc_overview}
\end{figure}

The proposed system consists of a VLC transmitter and a batteryless VLC-BLE receiver node designed for energy harvesting and indoor communication. The block diagram of the proposed system is presented in Figure \ref{fig:vlc_system}. 
The diagram illustrates the overall architecture of the system, including an energy harvester with a storage element, the analog front-end, and an MCU platform comprising a main processor and a dedicated low-power coprocessor. 
The transmitter uses a white LED based on phosphor to enable VLC transmission at frequencies above 800 Hz while maintaining standard indoor lighting conditions.

The detection of the VLC frame is coordinated with the open-circuit phase of the harvester. This timing reduces the switching noise on the sensing channel, enabling a more accurate sampling of the incoming optical signals. The VLC frame contains two parts, including a preamble, which is a fixed indicator of eights bits, and \SI{24}{bits} of the payload. During the open-circuit phase of the harvester, which lasts \SI{82}{\milli\second}, the system is able to receive a full data frame transmitted at an optical clock rate \SI{800}{\hertz}. Another portion inside the VLC frame is used for the sensing channel which allows for detecting the presence of VLC transmission, allowing the coprocessor to trigger an alert interrupt to the MCU and decide to run the main application or initiate VLC decode.

 The batteryless node uses a conventional solar panel, sized to make the device pocket-portable.  The analog front-end is designed to filter out the DC component of the solar panel voltage and condition the signal into a range suitable for detection by the MCU. 
Figure~\ref{fig:vlc_analog_front_end} shows the analog front-end circuit, which includes a high-pass filter with a cutoff frequency of approximately \SI{215}{\hertz}.
The energy harvester efficiently stores the energy collected from the solar panel in a capacitor and provides a stable output voltage to power the MCU. 
In this design, the capacitor voltage is monitored to ensure that either the main application or the VLC application is executed only when sufficient energy is available, thus preventing MCU shutdown due to excessive energy consumption. 
A voltage divider scales the capacitor voltage to a range compatible with the MCU’s ADC input.

Inside the MCU, two processors are used: a low-power coprocessor for continuous monitoring and a main processor for application execution. The low-power coprocessor is activated using the open-circuit indicator signal from the harvester chip to sampling the capacitor voltage.
If the capacitor voltage is high enough, the low-power processor samples the digital output signal from the analog front-end during the open-circuit sampling window of the harvester. Based on this sampled data, it decides whether to run the main application or the VLC application and then wakes up the main processor accordingly. 
The main processor runs either a main application, which transmits a broadcast BLE packet, or a VLC application, which decodes information from the VLC frame and then sends a BLE packet based on the received data, creating a VLC-BLE echo node.

\subsection{Implementation}\label{subsec3}

The hardware structure of the VLC-BLE node was implemented on a Texas Instruments LaunchPad CC1352P1 development board, which includes a dedicated low-power 8-bit coprocessor optimized for low power consumption and efficient peripheral operation, capable of running independently of the main processor.
The energy harvesting module was performed using an e-Peas AEM10941 chip with maximum power point tracking (MPPT) functionality. This harvester chip integrates the indicator signal when the panel is in open-circuit condition.  This open-circuit indicator signal is used to activate the low-power processor via an interrupt on a GPIO pin.

The harvester was configured to provide a stable output voltage of \SI{2.5}{\volt}. This output voltage is enabled when the capacitor voltage exceeds \SI{3.67}{\volt} and is disabled when the capacitor voltage falls below \SI{2.8}{\volt}. The maximum voltage of the capacitor is set to \SI{4.5}{\volt}. A 470$\mu$F capacitor was used as the storage element.
The analog front-end was implemented using a low-current operational amplifier MAX4007 to ensure minimal power consumption.
For the software, Zephyr RTOS was used to handle high-level application tasks, including VLC decoding, BLE communication, coprocessor initialization, and power management states. 

\section{Results}\label{sec4}

\begin{figure}[htbp]
  \centering
  \begin{minipage}[t]{0.38\linewidth}
    \centering
    \includegraphics[width=0.6\linewidth]{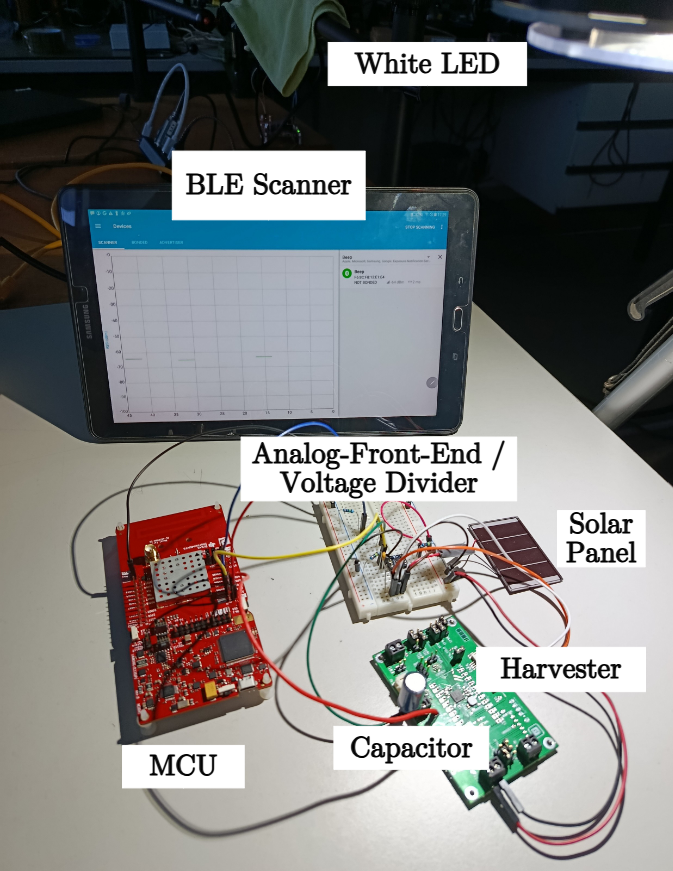}
    \caption{Experimental setup used to demonstrate the effective performance of the VLC-BLE application.}
    \label{fig:experimental_setup}
  \end{minipage}
  \hfill
  \begin{minipage}[t]{0.58\linewidth}
    \centering
    \includegraphics[width=0.9\linewidth]{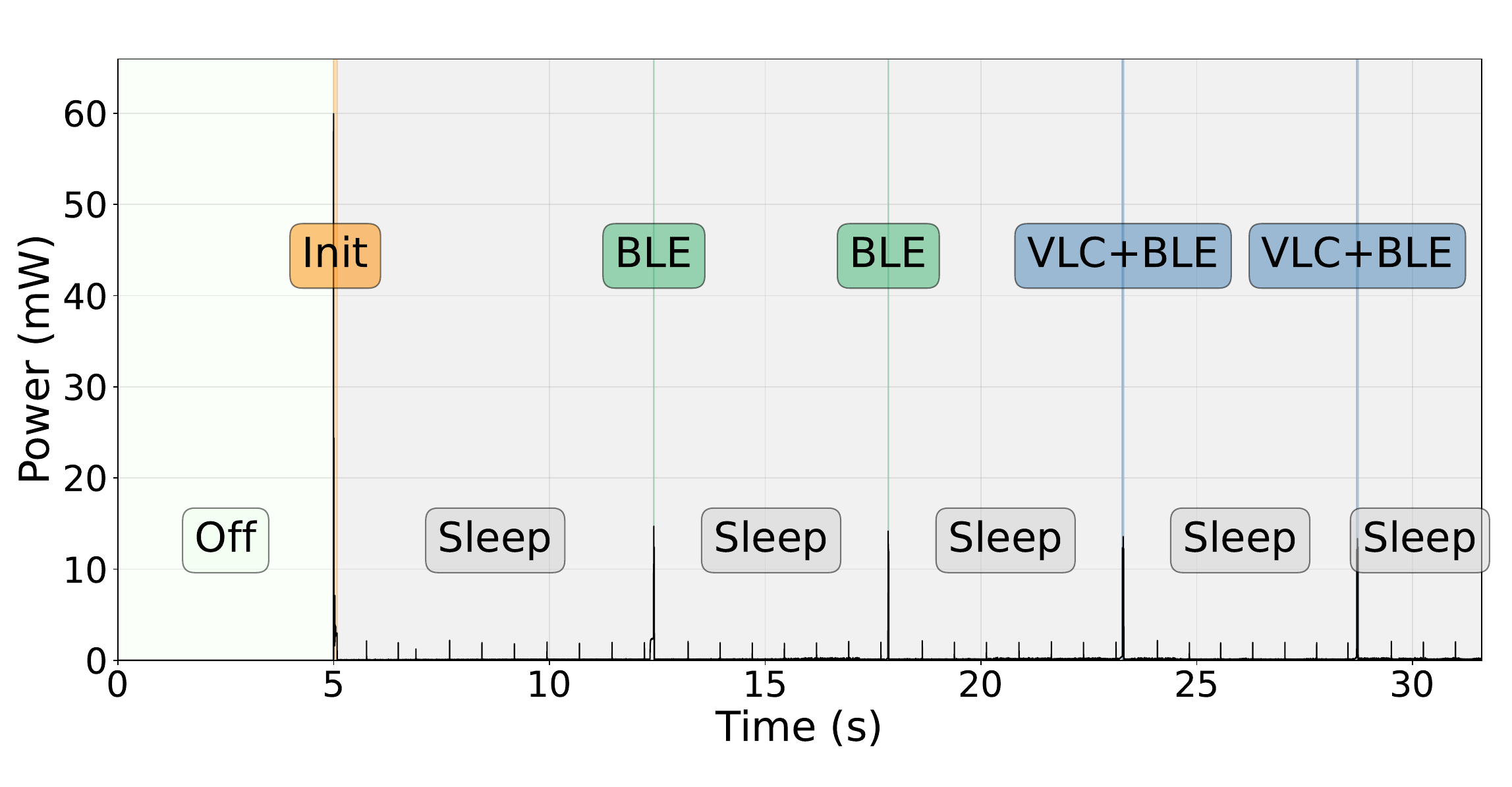}
    \caption{Power consumption profile of system phases including OFF, INIT, BLE, VLC+BLE, and SLEEP states.}
    \label{fig:power_profile}
  \end{minipage}
\end{figure}

The effectiveness of the proposed system was demonstrated using the experimental setup shown in Figure~\ref{fig:experimental_setup}. 
This figure displays the main components of the system: the light source for VLC transmission, the solar panel, the harvester evaluation kit with the storage element, and the MCU on an evaluation board. 
The figure also shows a tablet acting as a BLE receiver, displaying the BLE packets sent from the TI LaunchPad.  
Figure~\ref{fig:power_profile} presents a time-based power consumption profile of the MCU, illustrating the different operational states. 
The OFF state corresponds to when the MCU supply voltage is zero. Once the voltage supply is enabled, the INIT state begins, which includes the MCU startup and the execution of the main thread responsible for initializing the low-power coprocessor.  

The MCU then enters the SLEEP state, waiting for the capacitor voltage to exceed the overcharge threshold. When this occurs and the harvester activates the open-circuit signal, the system transitions to either the BLE or VLC+BLE state. 
Both states include the execution of a decision algorithm on the coprocessor, the wake-up of the main processor, and the execution of the corresponding application code.  
The execution of two consecutive BLE or VLC+BLE operations occurs at a minimum interval of 5\,s, a fixed parameter of the harvester chip. During this interval, the MCU remains in the SLEEP state, allowing the storage capacitor to recharge and reach the overcharge threshold again before the next event.  

The power, energy consumption, and execution time for each state were measured and are summarized in Table~\ref{tab:power_results_avg_std}. 
According to this characterization, the batteryless node consumes less than 0.030\,mW during the SLEEP state, which is sufficiently low to allow the capacitor to recharge in under 5\,s.  
Based on the harvester’s threshold configuration, the energy required to complete the INIT state, estimated at $0.345 \pm 0.033$\,mJ, is low enough to ensure that the system can reliably execute this state immediately after the voltage supply is activated, without discharging the capacitor below the threshold that would deactivate the supply. 
Similarly, the energy requirements for the BLE and VLC+BLE states are also low enough to allow full execution without triggering a shutdown due to insufficient energy.
Overall, the system exhibits low variability across all operational states, as evidenced by the small standard deviations in energy, power, and timing. This is especially notable in the BLE and VLC+BLE states, where the power and energy variations are minimal, indicating reliable behavior during wireless transmission.

\begin{table}[!t]
\caption{Energy, power consumption, and execution time characterization for each system state.
Here, $E_{\text{avg}}$, $P_{\text{avg}}$, and $t_{\text{avg}}$ denote the average energy, power, and execution time, respectively, while $E_{\text{std}}$, $P_{\text{std}}$, and $t_{\text{std}}$ represent their corresponding standard deviations.}
\label{tab:power_results_avg_std}
\begin{tabular*}{\textwidth}{@{\extracolsep\fill}|l|c|c|c|c|c|c|}
\hline
\textbf{State} &
$E_{\text{avg}}$ [mJ] & $E_{\text{std}}$ [mJ] &
$P_{\text{avg}}$ [mW] & $P_{\text{std}}$ [mW] &
$t_{\text{avg}}$ [s]  & $t_{\text{std}}$ [s]  \\
\hline
INIT     & 0.345 & 0.033 & 3.756 & 0.295 & 0.091 & 0.002 \\
SLEEP    & 0.151 & 0.024 & 0.030 & 0.004 & 5.017 & 0.842 \\
BLE      & 0.314 & 0.001 & 2.986 & 0.012 & 0.105 & 0     \\
VLC+BLE  & 0.345 & 0.002 & 2.925 & 0.011 & 0.118 & 0     \\
\hline
\end{tabular*}
\end{table}

\section{Conclusions}\label{sec5}
In this paper, we demonstrated the implementation of a batteryless system prototype that uses VLC to receive information from a light source, while simultaneously harvesting optical energy to power the MCU. By leveraging the open-circuit phase used by the harvester chip to sample the solar panel voltage, the system is able to capture and decode VLC signals with reduced interference, thereby simplifying the decoding process. The prototype is capable of reliably receiving a 32-bit VLC frame every 5\,s at a transmission rate of 800\,bps. The hardware and software architecture of the batteryless IoT node are designed to ensure low power consumption, maintaining sleep-state power below 30\,$\mu$W. Based on a sensing process implemented in the coprocessor, the system can autonomously decide whether to run a BLE application or a VLC+BLE application.  As future work, enhancing the slew rate of the operational amplifier used in the VLC analog front-end is identified as a high-priority improvement to support higher data rates. Additionally, exploring the use of VLC for firmware updates on the batteryless device is proposed as a promising extension of this work.

\bibliographystyle{splncs04}
\bibliography{references}

\end{document}